\documentclass{jfm}
\pdfoutput=1

\usepackage{graphicx}
\usepackage{natbib}
\usepackage{journals}
\usepackage{amsmath}
\usepackage{upmath}
\usepackage{amssymb}
\usepackage{longtable}
\usepackage{gensymb}
\usepackage{rotating}
\usepackage{multirow}
\usepackage{amsbsy}
\usepackage[colorlinks=True,citecolor=blue,linkcolor=blue,%
			urlcolor=blue]{hyperref}

\usepackage{xcolor}

\def\vec#1{\ensuremath{\mathchoice{\mbox{\boldmath$\displaystyle#1$}}
{\mbox{\boldmath$\textstyle#1$}}
{\mbox{\boldmath$\scriptstyle#1$}}
{\mbox{\boldmath$\scriptscriptstyle#1$}}}}

\newsavebox{\astrutbox}
\sbox{\astrutbox}{\rule[-5pt]{0pt}{20pt}}

\title{Latitudinal regionalization of rotating spherical shell convection}

\author[T. Gastine and J.~M. Aurnou]%
{Thomas Gastine$^1$%
  \thanks{Email address for correspondence: gastine@ipgp.fr},\ns Jonathan M.  
 Aurnou$^2$}

\affiliation{$^1$Universit\'e Paris Cit\'e, Institut de Physique du Globe de 
Paris,  UMR 7154 CNRS, 1 rue Jussieu, F-75005 Paris, France, \\ $^2$Department 
of Earth, Planetary, and Space Sciences, University of California, Los Angeles, 
CA 90095, USA.}

\pubyear{2022}
\volume{650}
\pagerange{119--126}
\date{?; revised ?; accepted ?. - To be entered by editorial office}
\begin{document}

\maketitle

%
%
\begin{abstract}
Convection occurs ubiquitously on and in rotating geophysical and astrophysical 
bodies. Prior spherical shell studies have shown that the convection dynamics in 
polar regions can differ significantly from the lower latitude, equatorial 
dynamics. Yet most spherical shell convective scaling laws use globally-averaged 
quantities that erase latitudinal differences in the physics. Here we quantify 
those latitudinal differences by analyzing spherical shell simulations in terms 
of their regionalized convective heat transfer properties. This is done by 
measuring local Nusselt numbers in 
two specific, latitudinally separate, portions of the shell,
the polar and the equatorial regions, $Nu_p$ and $Nu_e$, respectively. In 
rotating spherical shells, convection first sets in outside the tangent cylinder 
such that equatorial heat transfer dominates at small and moderate 
supercriticalities. We show that the buoyancy forcing, parameterized by the 
Rayleigh number $Ra$, must exceed the critical equatorial forcing by a factor of 
$\approx 20$ to trigger polar convection within the tangent cylinder. Once 
triggered, $Nu_p$ increases with $Ra$ much faster than does $Nu_e$. The 
equatorial and polar heat fluxes then tend to become comparable at 
sufficiently high $Ra$. Comparisons between the polar convection data 
and Cartesian numerical simulations reveal quantitative agreement between the 
two geometries in terms of heat transfer and averaged bulk temperature 
gradient. This agreement indicates that spherical shell rotating convection 
dynamics are accessible both through spherical simulations and via reduced 
investigatory pathways, be they theoretical, numerical or experimental. 
\end{abstract}


\section{Introduction}
It has long been known that spherical shell rotating convection significantly differs between the low latitudes 
\citep[e.g.,][]{Busse77, Gillet06} situated outside the 
axially-aligned cylinder that circumscribes the inner spherical 
shell boundary  (the tangent cylinder, TC)
and the higher latitude polar regions lying within the TC
\citep[e.g.,][]{Aurnou03, Sreenivasan06, Aujogue18, Cao18}.
Further, in the atmosphere-ocean literature, latitudinal separation into polar, 
mid-latitude, extra-tropical and tropical zones is essential to accurately model 
the large-scale dynamics \cite[e.g.,][]{Vallis17}.  
Yet few scaling studies of spherical shell convection consider the innate 
regionalization of the dynamics \cite[cf.][]{Wang21}, and instead
mostly focus on globally-averaged quantities \citep[e.g.,][]{Gastine16,Long20}.

In the turbulent rapidly-rotating limit, theory requires the convective heat 
transport to be independent of the fluid diffusivities irregardless of system 
geometry. This yields \citep[e.g.][]{Julien12, Plumley19}
\begin{equation}
Nu \sim (Ra/Ra_c)^{3/2} \sim \widetilde{Ra}^{3/2}Pr^{-1/2} \sim Ra^{3/2} E^{2} 
Pr^{-1/2}\,,
 \label{eq:ult}
\end{equation}
where, defined explicitly below, the Nusselt number $Nu$ is the nondimensional 
heat transfer, $Ra$ ($Ra_c$) denotes the (critical) Rayleigh number, $E$ is the 
Ekman number, $Pr$ is the Prandtl number, and $\widetilde{Ra} \equiv 
Ra\,E^{4/3}$ expresses the generalized convective supercriticality 
\citep{Julien12}.

Cylindrical laboratory experiments with $Pr\approx 7$ and 
Cartesian (planar) numerical simulations with $Pr=(1, 7)$ and no-slip boundaries 
with 
$Ra/Ra_c \lesssim 10$ reveal a steep scaling $Nu \sim (Ra/Ra_c)^\beta$ 
with $\beta\approx 3$ \citep{King12, Cheng15, Cheng18}. By comparing numerical 
models with stress-free and no-slip boundaries, \citet{Stellmach14} showed that 
the steep $\beta\approx 3$ scaling is an Ekman pumping effect 
\citep[cf.][]{Julien16}. 
For larger supercriticalities, $\beta$ decreases and
gradually approaches (\ref{eq:ult}).
This $\beta \approx 3$ regime is expected to hold as long as the thermal 
boundary layers 
are in quasi-geostrophic balance, a condition approximated by $Ra\,E^{8/5} 
\lesssim 1$ \citep{Julien12a}.

Globally-averaged quantities in spherical shell models present several 
differences with the planar configuration. In particular, no 
steep $\beta \approx 3$ exponent is observed. \citet{Gastine16} showed that the 
globally-averaged heat transfer first follows a $Nu-1 \sim Ra/Ra_c -1$ 
weakly-nonlinear scaling for $Ra \leq 6\,Ra_c$ before transitioning to 
a scaling close to (\ref{eq:ult}) for $Ra > 6\,Ra_c $ and $Ra E^{8/5} < 0.4$. 
Spherical shell models with a radius ratio $r_i/r_o=0.35$ and 
fixed-flux thermal 
conditions recover similar global scaling behaviors,
though with a slightly larger exponent $\beta \approx 1.75$ for $E=2\times 
10^{-6}$ \citep{Long20}. Because the Ekman pumping enhancement of heat 
transfer is maximized when rotation and gravity are aligned, $\beta$ is lower 
in the equatorial regions of spherical shells. This explains why 
globally-averaged spherical $\beta$ values cannot attain the $\beta \approx 3$ 
values found in planar (polar-like) studies. 

Recently, \citet{Wang21} analysed heat transfer 
within the equatorial regions, at mid-latitudes, and inside the entire 
TC. They argued that the mid-latitude scaling in their 
models, similar to \citet{Gastine16}'s global scaling, follows the 
diffusion-free scaling (\ref{eq:ult}), whilst the region inside the TC follows 
a $\beta \approx 2.1$ trend.
This TC scaling exponent is significantly smaller than those obtained in planar 
models, possibly because of the finite inclination angle between gravity and 
the 
rotation axis averaged over the volume of the TC.

Following \citet{Wang21}, this study aims to better characterize the 
latitudinal variations in rotating convection dynamics and quantify the 
differences between spherical and non-spherical geometries. To do so, we carry 
out local heat transfer analyses in the polar and equatorial regions over an 
ensemble of $Pr=1$ rotating spherical shell 
simulations with $r_i/r_o=0.35$ and $r_i/r_o=0.6$.

\section{Hydrodynamical model}
\label{sec:model}

We consider a volume of fluid bounded by two spherical surfaces of inner radius 
$r_i$ and outer radius $r_o$ rotating about the $z$-axis with a constant 
rotation rate $\Omega$. Both boundaries are mechanically no-slip and are held at constant temperatures 
$T_o=T(r_o)$ and $T_i=T(r_i)$. 
We adopt a dimensionless formulation 
of the Navier-Stokes equations using the shell gap $d=r_o-r_i$ as the reference 
lengthscale, the temperature contrast $\Delta T=T_o-T_i$ as the temperature 
unit, and the inverse of the rotation rate $\Omega^{-1}$ as the time scale. 
Under the Boussineq approximation, this yields the following set of 
dimensionless equations for the velocity $\vec{u}$ and temperature $T$ 
expressed in spherical coordinates
\begin{equation}
 \dfrac{\partial \vec{u}}{\partial 
t}+\vec{u}\cdot\vec{\nabla}\vec{u}+2\vec{e_z}\times\vec{u}=-\vec{\nabla}p + 
\dfrac{Ra E^2}{Pr} T\,g(r)\vec{e_r}+ E\,\vec{\nabla}^2\vec{u}, 
\quad\vec{\nabla}\cdot\vec{u}=0,
\label{eq:ns}
\end{equation}
\begin{equation}
 \dfrac{\partial T}{\partial t}+\vec{u}\cdot\vec{\nabla}T = 
\dfrac{E}{Pr}\vec{\nabla}^2 T\,,
\label{eq:temp}
\end{equation}
where $p$ corresponds to the non-hydrostatic pressure, $g$ to gravity and 
$\vec{e_r}$ 
($\vec{e_z}$) denotes the unit vector in the radial (axial) direction.
The above equations 
are governed by the dimensionless Rayleigh, Ekman and Prandtl numbers, 
respectively defined by
\begin{equation}
 Ra=\dfrac{\alpha g_o \Delta T d^3}{\nu \kappa},\ E=\dfrac{\nu}{\Omega d^2},\ 
 Pr = \dfrac{\nu}{\kappa},
 \label{eq:numbers}
\end{equation}
where $\nu$ and $\kappa$ correspond to the constant kinematic viscosity and 
thermal diffusivity, and $\alpha$ is the thermal expansion coefficient.
Two spherical shell configurations are employed: (\textit{i}) a 
thin shell with $r_i/r_o=0.6$ under the assumption of a centrally-condensed 
mass with $g=(r_o/r)^2$ \citep{Glatz1}; (\textit{ii}) a 
self-gravitating thicker spherical shell model with $r_i/r_o=0.35$ and 
$g=r/r_o$. The latter corresponds to the standard configuration employed in 
numerical models of Earth's dynamo \citep[e.g.][]{Christensen06, Schwaiger19}. 
We consider numerical simulations with $10^4 \leq Ra \leq 10^{11}$, $10^{-7} 
\leq E \leq 10^{-2}$ and $Pr=1$ computed with the open source code 
\texttt{MagIC}\footnote{\url{https://github.com/magic-sph/magic}} 
\citep{Wicht02,Gastine12}. We mostly build the current study on existing 
numerical simulations from \cite{Gastine16} and \cite{Schwaiger21} and continue 
their time integration to gather additional diagnostics when required.

In the following analyses overbars denote time averages, triangular brackets denote
azimuthal averages and square brackets denote averages about the angular sectors 
comprised between the colatitudes $\theta_0-\alpha$ and $\theta_0+\alpha$ in 
radians:
\[
\bar{f} = \int_{t_0}^{t_0+\tau}f\mathrm{d}t,\quad
 \langle f \rangle = \dfrac{1}{2\pi}\int_{0}^{2\pi} 
f(r,\theta,\phi,t)\mathrm{d}\phi, \quad
 \left[ f \right]_{\theta_0}^{\alpha} = 
\dfrac{1}{\mathcal{S}_{\theta_0}^\alpha}\int_{\mathcal{S}_{\theta_0}^\alpha}
 f(r,\theta,\phi,t)\mathrm{d}\mathcal{S},
\]
with $\mathrm{d}\mathcal{S} = \sin\theta \mathrm{d}\theta$ 
and 
$\mathcal{S}_{\theta_0}^\alpha=\int_{\min(\theta_0-\alpha,0)}^{ 
\max(\theta_0+\alpha,\pi) } \sin\theta\mathrm{d}\theta$.

For the sake of clarity, we introduce the following notations to characterize 
the time-averaged  radial distribution of temperature
\[
 \vartheta(r)=[\langle  \bar{T} 
\rangle ]_{\pi/2}^{\pi/2}, \quad
 \vartheta_e(r)= [\langle  \bar{T} 
\rangle ]_{\pi/2}^{\pi/36}, \quad
 \vartheta_p(r)=\dfrac{1}{2}\left([\langle  \bar{T} 
\rangle ]_{0}^{\pi/36}+[\langle  \bar{T} 
\rangle ]_{\pi}^{\pi/36}\right)\,,
\]
where $\vartheta_e$ and $\vartheta_p$ correspond to the averaged radial 
distribution of temperature in the equatorial and polar regions,
respectively, and $\alpha = \pi/36$ rad corresponds to $5^\circ$ in 
colatitudinal angle.
The schematic shown in Fig.~\ref{fig:th_Ranu}(\textit{a}) highlights the 
fluid volumes involved in these measures.
The value of $\alpha=5^\circ$ is quite arbitrary and has been adopted to allow 
a comparison of polar data with local planar Rayleigh-B\'enard 
convection (hereafter RBC) models while keeping a
sufficient sampling. 

To quantify the differences between the heat transfer 
in the polar and equatorial regions, we introduce a Nusselt 
number that depends on colatitude $\theta$ via
\begin{equation}
 Nu_i(\theta) = \dfrac{\left.\frac{\mathrm{d} \langle \bar{T} \rangle 
}{\mathrm{d}{r}}\right|_{r_i}}{
\left.\frac{\mathrm{d} T_c}{\mathrm{d}{r}}\right|_{r_i}}, \quad
Nu_o(\theta) = \dfrac{\left.\frac{\mathrm{d} \langle \bar{T} \rangle 
}{\mathrm{d}{r}}\right|_{r_o}}{
\left.\frac{\mathrm{d} T_c}{\mathrm{d}{r}}\right|_{r_o}}, \quad
\dfrac{\mathrm{d} T_c}{\mathrm{d} r} = -\dfrac{r_i r_o}{r^2}\,,
\end{equation}
where $T_c$ corresponds to the dimensionless temperature of the conducting 
state.
The corresponding local Nusselt numbers in the equatorial and polar regions are 
then defined by
\begin{equation}
 Nu_e=[Nu(\theta)]_{\pi/2}^{\pi/36}, \quad
 Nu_p=\dfrac{1}{2}\left([\langle  Nu(\theta)
\rangle ]_{0}^{\pi/36}+[\langle  Nu(\theta)
\rangle ]_{\pi}^{\pi/36}\right)\,.
 \label{eq:nuloc}
\end{equation}
We finally introduce the mid-shell time-averaged temperature gradient in the 
polar region
\begin{equation}
 \partial T = \dfrac{%
-\left.\frac{\mathrm{d}\vartheta_p}{
\mathrm {d} r}\right|_{r=r_m}}{%
-\left.\frac{\mathrm{d}T_c}{
\mathrm {d} r}\right|_{r=r_m}
}, \quad r_m=\dfrac{1}{2}(r_i+r_o)\,,
\label{eq:beta}
\end{equation}
where normalisation by the conductive temperature gradient allows us to 
compare the scaling behaviour of $\partial T$ between spherical 
shells of different radius ratio values, $r_i/r_o$, and planar models.

\section{Results}
\label{sec:results}



\begin{figure}
 \centering
 \includegraphics[width=\textwidth]{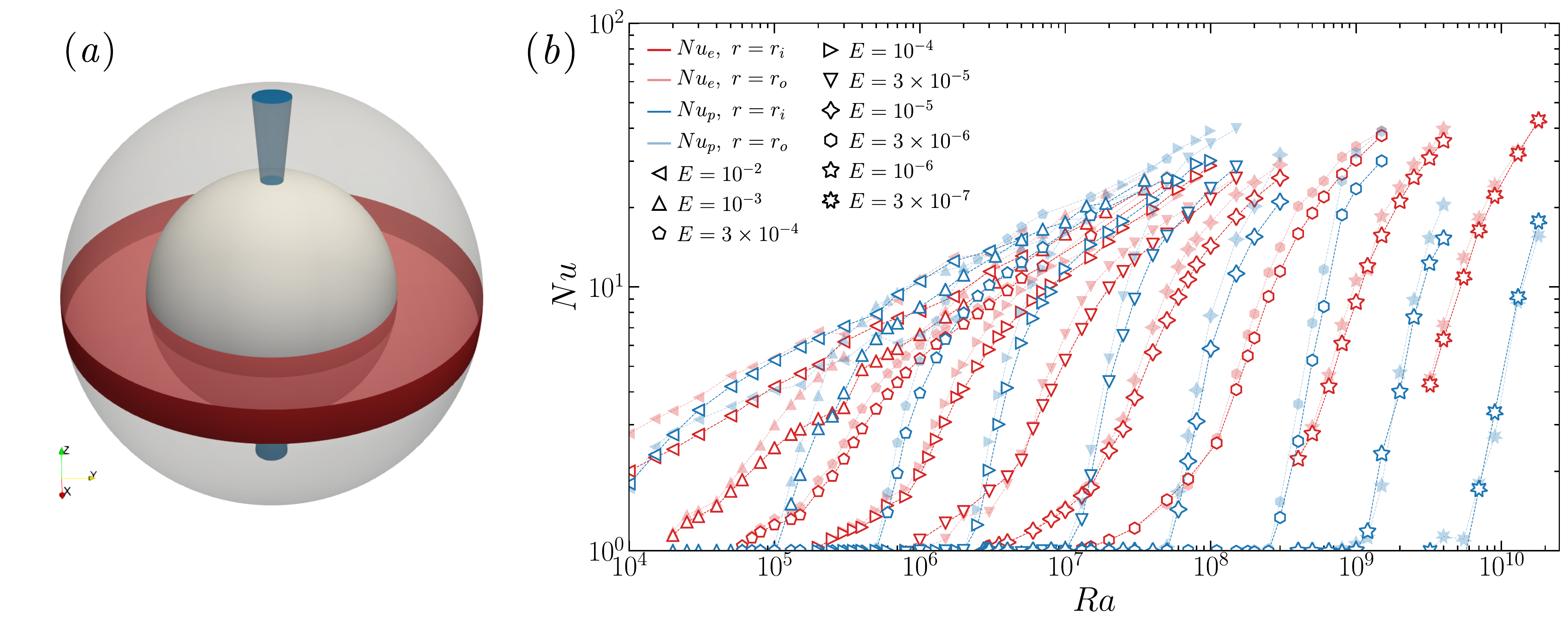}
 \caption{(\textit{a}) Schematic showing the area selection to compute 
(\ref{eq:nuloc}), the local polar (blue) and equatorial (red) Nusselt numbers.  
(\textit{b}) Time-averaged local Nusselt numbers in the polar ($Nu_p$) and 
equatorial ($Nu_e$) regions as a function of the Rayleigh number for spherical 
shell simulations with $r_i/r_o=0.6$ and $g=(r_o/r)^2$ and $Pr=1$ 
\citep{Gastine16}. The different Ekman numbers are denoted by different symbol 
shapes, the two spherical shells surfaces $r_i$ and $r_o$ are marked by open and 
filled symbols, and by lower levels of opacity, respectively.}
 \label{fig:th_Ranu}
\end{figure}

Figure~\ref{fig:th_Ranu}(\textit{b}) shows $Nu_p$ and $Nu_e$ as a 
function of $Ra$ for various $E$ at both boundaries, $r_i$ and $r_o$, for 
spherical shell simulations with $r_i/r_o=0.6$ and $g=(r_o/r)^2$. Rotation 
delays the onset of convection such that the critical Rayleigh number required to trigger 
convective motions increases with decreasing Ekman number, $Ra_c \sim 
E^{-4/3}$.
Convection first sets in outside the tangent cylinder
\citep[e.g.][]{Dormy04}. For each Ekman number, heat transfer 
behaviour in the equatorial regions (red symbols) first raises slowly following a weakly nonlinear scaling 
 \citep[e.g.][]{Gillet06}, before gradually rising in the vicinity of $Nu_e \approx 
2$. At $Nu_e \gtrsim 2$, the heat transfer increases more steeply with $Ra$, before 
gradually tapering off toward the non-rotating RBC trend 
\citep[e.g.][]{Gastine15}. 
For $Ra/Ra_c > \mathcal{O}(10)$, convection sets in the polar regions and $Nu_p$ 
steeply rises with $Ra$ with a much larger exponent than $Nu_e$. At still larger 
forcings, the slope of $Nu_p$ gradually decreases and comparable amplitudes 
in polar and equatorial heat transfers are observed.
Heat transfer scalings at both spherical shell boundaries $r_i$ and $r_o$ 
follow similar trends.

\begin{figure}
 \centering
 \includegraphics[width=\textwidth]{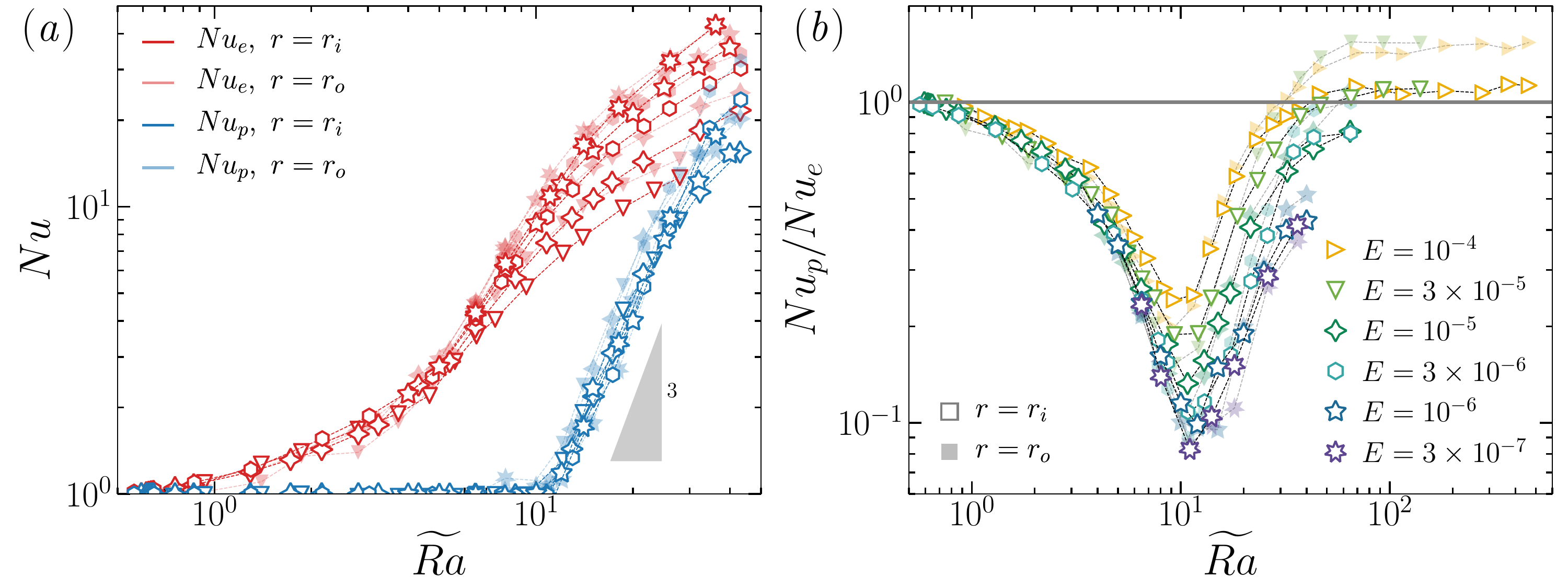}
 \caption{(\textit{a}) Nusselt number in the polar ($Nu_p$) and in the 
equatorial ($Nu_e$) regions as a function of $\widetilde{Ra}=Ra\,E^{4/3}$ 
in the $r_i/r_o = 0.6$ simulations.
The symbols carry the same meaning as in 
Fig.~\ref{fig:th_Ranu} but with only the $Ra\,E^{8/5} < 2$ simulations retained. (\textit{b}) Ratio of polar and equatorial heat 
transfer $Nu_p/Nu_e$ as a function of $\widetilde{Ra}$ for both spherical shell 
boundaries and $E \leq 10^{-4}$.}
 \label{fig:thRaEkNu}
\end{figure}

Figure~\ref{fig:thRaEkNu} shows (\textit{a}) $Nu_p$ and $Nu_e$ and (\textit{b}) 
their ratio $Nu_p/Nu_e$ plotted at both boundaries as a function of the 
supercriticality parameter $\widetilde{Ra} = Ra E^{4/3}$. For $\widetilde{Ra} < 
4$, $Nu_e$ increases following the weakly nonlinear form $Nu_e-1\sim Ra/Ra_c-1$ 
\citep[][\S3.1]{Gastine16}. For larger supercriticalities, the $Nu_e$ 
scaling steepens and an additional $E$-dependence causes the data to 
fan out, possibly because these highest $\widetilde{Ra}$ cases do not fulfill 
$Ra\,E^{8/5}<0.4$. 
There is no clear power law scaling in the $Nu_e (\widetilde{Ra} < 10)$ data, 
but the steepest local slope yields $\max(\beta) \approx 1.9$ in the $5 \leq 
\widetilde{Ra} \leq 10$ range.

Best fits to the Fig.~\ref{fig:thRaEkNu}(\textit{a}) data show that polar 
convection onsets at $\widetilde{Ra} (E) = 11.2 \pm 0.3$ in the $r_i/r_o = 0.6$ 
simulations.  The mean value of the critical polar Rayleigh number is 
\begin{equation}
Ra_c^p = 11.2\,E^{-4/3}.
\label{RaP}
\end{equation} 
Although the polar onset of convection, estimated via $Ra_c^p \, E^{4/3}$, 
remains nearly constant, the global (e.g., low latitude) onset value, estimated 
by $Ra_c \, E^{4/3}$, varies by a factor of $\approx 2$ over our $E$ range. 
Their ratio then yields 
\begin{equation}
Ra_c^p (E)/Ra_c(E) = 20 \pm 5. 
\end{equation}
This means that rotating convection does not typically onset in the polar 
regions until the lower latitude convection is already 20 times supercritical 
and is already operating under highly supercritical conditions. This difference 
in equator versus polar convective onsets imparts a significant regionalization 
to spherical shell rotating convection right from the get go.

We find, throughout this investigation, that polar rotating convection compares 
closely to its plane layer counterpart. However, it is not expected that the 
polar critical Rayleigh number will exactly agree with plane layer 
predictions, due to the effects of finite spherical curvature as well as the 
radial variations of gravity in these $r_i/r_o = 0.6$ simulations. In the 
rapidly-rotating thin shell limit, in which $r_i/r_o \rightarrow 1$ and $E$ is 
kept asymptotically small, $Ra_c^p$ will likely approach the 
planar value.
Still, the polar scaling in \eqref{RaP} is found to be 51\% of the plane layer 
$E \rightarrow 0$ scaling prediction, $Ra_c = 21.9\,E^{-4/3}$ 
\citep[][]{Kunnen21}, and to be 56\% of \citet{Niiler65}'s finite Ekman number, 
no-slip plane layer $Ra_c $ prediction at $E = 10^{-6}$. In addition to the 
similarity in critical $Ra$ values, it is found that the polar heat transfer 
$Nu_p$ rises sharply once polar convection onsets, following a $Nu_p \sim 
\widetilde{Ra}^3$ scaling that  
matches the heat transfer scalings found in no-slip planar simulations carried 
out over the same $(E, Pr)$ ranges \citep{King12, Stellmach14, Aurnou15}.

Figure~\ref{fig:thRaEkNu}(\textit{b}) shows the ratio of polar to equatorial 
heat transport, 
which follows a distinct v-shape trend that can be decomposed in three regions. 
(\textit{i}) For $\widetilde{Ra}< 11.2$, $Nu_p\approx 1$ and the ratio depends 
directly on $Nu_e=f(\widetilde{Ra})$. 
(\textit{ii}) For $11.2 < \widetilde{Ra} \lesssim 30$, $Nu_p$ raises much faster than 
$Nu_e$ hence increasing $Nu_p/Nu_e$. (\textit{iii}) When rotational 
effects become less influential, $Nu_p/Nu_e \approx 1$ at
$r_i$ and $Nu_p/Nu_e\approx 1.5$ at $r_o$.

\begin{figure}
 \centering
 \includegraphics[width=\textwidth]{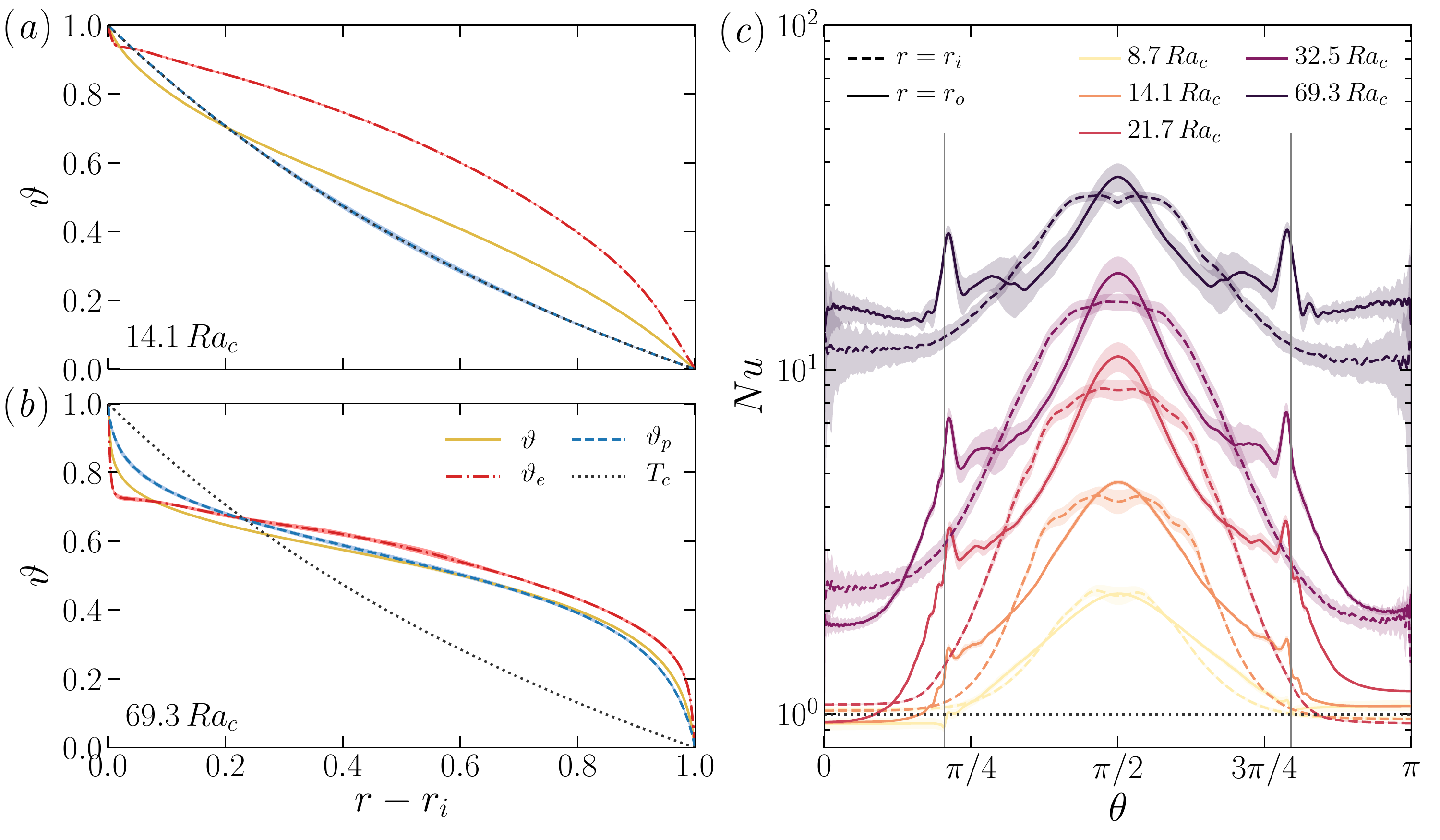}
 \caption{(\textit{a-b}) Radial profiles of time-averaged temperature in 
the polar regions (blue dashed line), in the equatorial region (red dot-dashed line) and 
averaged of the entire spherical surface (tan solid line). For comparison, the 
conducting temperature profile $T_c$ is also plotted as a black dotted line. Panel 
(\textit{a}) corresponds to $r_i/r_o=0.6$, $g=(r_o/r)^2$, $E=10^{-6}$, 
$Ra=6.5\times 10^8$, $Pr=1$, while panel (\textit{b}) corresponds to
$r_i/r_o=0.6$, $g=(r_o/r)^2$, $E=10^{-6}$, $Ra=3.2\times 10^9$ and $Pr=1$.
(\textit{c}) Time-averaged local Nusselt number at both spherical shell 
boundaries as a function of the colatitude for simulations with $r_i/r_o=0.6$, 
$g=(r_o/r)^2$, $E=10^{-6}$, $Pr=1$ and increasing supercriticalities. Solid 
(dashed) lines correspond to $r_i$ ($r_o$). The vertical solid lines mark the 
location of the tangent cylinder. In all panels, the shaded regions correspond 
to one standard deviation about the time averages.}
 \label{fig:profiles}
\end{figure}


Figure~\ref{fig:profiles}(\textit{a}-\textit{b}) shows the time-averaged 
temperature profiles in the polar and equatorial regions ($\vartheta_p$ dashed 
lines and $\vartheta_e$ dot-dashed lines) alongside the volume-averaged 
temperature ($\vartheta$, solid line) for two numerical models with  
$r_i/r_o=0.6$, $g=(r_o/r)^2$, $E=10^{-6}$ and different $Ra$. For the case with 
$Ra\approx 14.1\,Ra_c$ (panel \textit{a}), low latitude convection is
active but has yet to start within the TC.
The mean temperature in the polar regions $\vartheta_p$ thus closely follows 
the conductive profile $T_c$ (dotted line), while in the equatorial region 
we observe the formation of a thin thermal boundary layer at $r_i$ and a 
decrease of the temperature gradient in the fluid bulk.
At larger convective forcing ($Ra\approx 69.3\,Ra_c$, pabel \textit{b}), 
convection is space-filling. The temperature profiles in the polar and 
equatorial regions become comparable and 
a larger fraction of the temperature contrast is accomodated in the 
thermal boundary layers.

Figure~\ref{fig:profiles}(\textit{c}) shows the latitudinal variations of the 
heat flux at both spherical shell boundaries for increasing supercriticalities.
These profiles confirm that convection first sets in outside the TC whilst the 
high-latitude regions remain close to the conductive $Nu = 1$ state up to 
$Ra_c^p$, and that the $Ra > Ra_c^p$ polar 
transfer rises quickly, thus reducing the latitudinal $Nu$ contrast.
Both spherical shell boundaries feature similar global 
trends, with interesting regionalized differences. The 
tangent cylinder (solid vertical lines) is visible, for instance, in the outer 
boundary heat transfer $Nu_o(\theta)$, manifesting itself in local maxima that 
persist between $15\,Ra_c$ and $70\, Ra_c$.

\begin{figure}
 \centering
 \includegraphics[width=\textwidth]{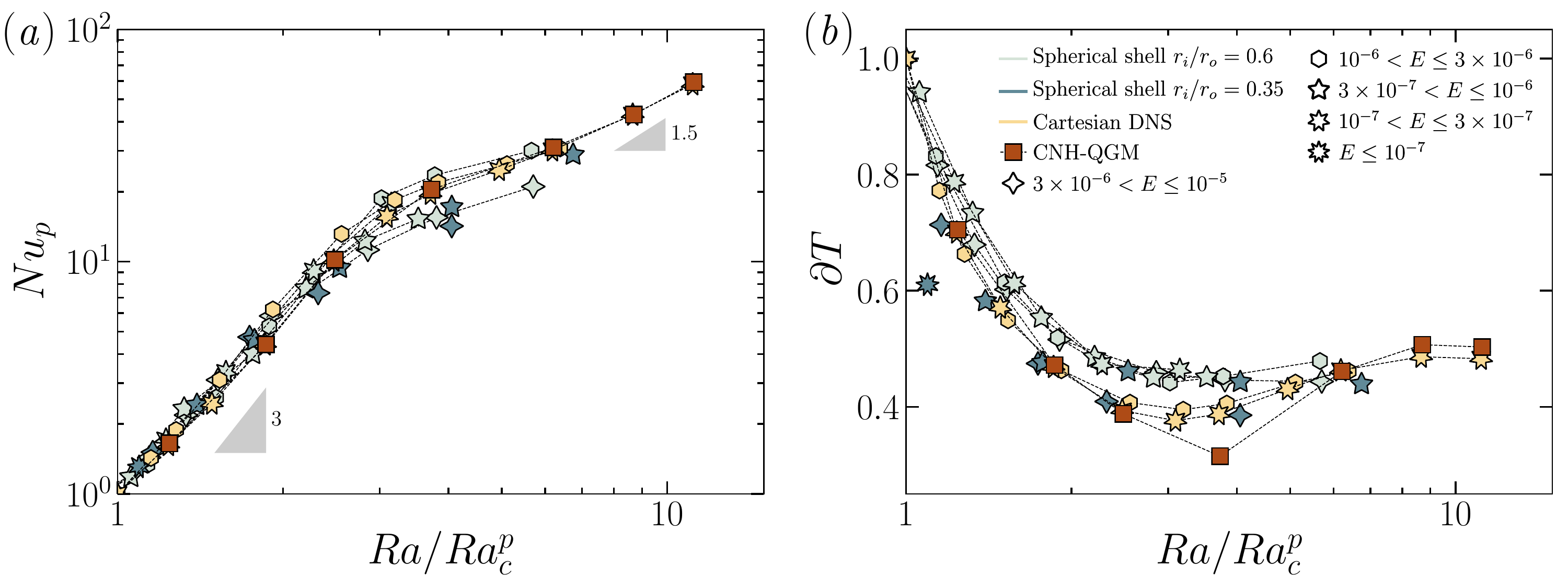}
 \caption{(\textit{a}) Nusselt number in the polar regions $Nu_p$ as a function 
of the local supercriticality $Ra/Ra_c^p$. (\textit{b}) Normalised 
mid-depth temperature gradient (Eq.~\ref{eq:beta}) in the polar regions 
$\partial T$ 
as a function of the local supercriticality. Spherical shell simulations 
include two configurations with $r_i/r_o=0.6$ and $g=(r_o/r)^2$ \citep[light 
blue symbols, from][]{Gastine16} and $r_i/r_o=0.35$ and $g=r/r_o$ 
\citep[dark blue symbols, from][]{Schwaiger21}. All the simulations with $E 
\leq 10^{-5}$ and $Nu_p > 1$ have been retained. Direct numerical simulations 
(DNS) in Cartesian geometry with periodic horizontal boundary conditions (light 
yellow symbols) come from \cite{Stellmach14}, while non-hydrostatic 
quasi-geostrophic models (CNH-QGM) (red symbols) come from \cite{Plumley16}.}
 \label{fig:NuPo}
\end{figure}

Figure~\ref{fig:NuPo} shows (\textit{a}) $Nu_p$ and (\textit{b}) 
normalized mid-depth polar temperature gradients $\partial T$ as a function of 
$Ra/Ra_c^p$ 
for spherical shell simulations with $r_i/r_o=0.6$ and $r_i/r_o=0.35$, 
and for Cartesian asymptotically reduced models \citep[e.g.,][]{Plumley16} and 
$E\geq 2\times 10^{-7}$, $Pr=1$ direct numerical simulations 
\citep{Stellmach14}. In this figure, $Ra_c^p$ is used for the critical $Ra$ 
values for spherical shell data, whereas standard planar $Ra_c$ values are used 
for the plane layer data. Good quantitative agreement is found in the $Nu_p$ and 
$\partial T$ data from spherical shell and 
planar models, with all the data sets effectively overlying one another. The $1 
\lesssim Ra/Ra_c^p \lesssim 3$ heat transfer follows a $Nu_p \sim (Ra/Ra_c)^3$ 
scaling in all the data sets. At larger supercriticalities, the scaling 
exponent 
of $Nu_p$ decreases and the 
asymptotic $\beta = 3/2$ scaling 
appears to be approached in the highest supercriticality planar cases.
The mid-depth temperature gradients quantitatively agree in all models as well, 
attaining a relatively large minimum value, $\partial T \approx 0.5$ near $Ra 
\approx 3\,Ra_c^p$, before increasing slightly in the highest supercriticality 
planar models. 

\section{Discussion}
\label{sec:conclu}

Globally-averaged heat transfer scalings for rotating convection differ between 
spherical and 
planar geometries with the latter yielding steeper $Nu$-$Ra$ scaling trends. 
By introducing regionalized measures of heat 
transfer, we have shown that this steep scaling can also be recovered in the 
polar regions of spherical shells. The comparisons in Fig.~\ref{fig:NuPo} 
reveals an almost perfect overlap in heat transfer data between the two 
geometries. Importantly, this demonstrates that local, non-spherical models can 
be used to understand spherical systems \citep[e.g.,][]{Julien12, Horn15, 
Cabanes17, Calkins18, Cheng18, Miquel18, Gastine19}.

Our regional analysis shows that the use of global volume-averaged properties 
to interpret spherical shell rotating convection can be misleading since such 
averages are often made over regions with significantly differing convection 
dynamics \citep[e.g.,][in rotating cylinders]{Ecke14,Lu21,Grannan22}. As such, 
it is quite likely that globally-averaged $\beta$ depends on the 
spherical shell radius ratio, $r_i/r_o$.  In higher $r_i/r_o$ shells, more of 
the fluid will lie 
within the TC and the globally-averaged $\beta$ will tend towards a polar value 
near $3$.  In contrast, lower $r_i/r_o$ shells should trend towards regional 
$\beta$ values below $2$, as found in our $Nu_e$ data.  We hypothesize further 
that the mid latitude $\beta \simeq 3/2$ scaling in \citep{Wang21} may 
represent a combination of the low and high latitude scalings, which could also 
be tested by varying $r_i/r_o$.

A similar argument may also explain \cite{Wang21}'s higher latitude, tangent 
cylinder heat transfer scaling  of $\beta = 2.1$. We postulate that 
measuring the rotating heat transfer away from the poles will always yield 
$\beta < 3$.  This may be further exacerbated if the heat transfer is measured 
across the tangent cylinder, which likely acts as a radial transport barrier 
\cite[e.g.,][]{Guervilly17, Cao18}. Thus, \citet{Wang21}'s $\beta \approx 2.1$ 
value may arise because their whole tangent cylinder measurements extend to far 
lower latitudes in comparison to the far tighter, pole-adjacent $Nu_p$ 
measurements made here that yield $\beta \approx 3$.

The polar heat transfer data in Figure \ref{fig:thRaEkNu} demonstrates a sharp 
convective onset value, with $Ra_c^p = (11.2 \pm 0.3)E^{-4/3}$ over our range 
of $r_i/r_o = 0.6$ models and $Ra_c^p / Ra_c = 20 \pm 5$.  It is likely that 
convective turbulence is space-filling in planetary fluid layers.  We argue 
then that realistic geophysical and astrophysical models of rotating convection 
require $Ra > Ra_c^p$. If the convection is rapidly-rotating as well, this 
constrains the convective Rossby number $Ro_{conv} = (Ra E^2 / Pr)^{1/2} 
\lesssim 0.1$ \citep[e.g.,][]{Christensen06, Aurnou20}.  Thus, space-filling 
rotating convective turbulence simultaneously requires $Ra \gtrsim 10 Ra_c^p$ 
and $Ro_{conv} \lesssim 1/10$, which then constrains that $E \lesssim 10^{-6}$ 
in $Pr \simeq 1$ models.  Such dynamical constraints are important for building 
accurate models of $Nu(\theta)$, which are essential to our interpretations of 
planetary and astrophysical observations. For instance, on the icy satellites, 
latitudinal changes in ice shell thickness and surface terrain likely reflect 
the latitudinally-varying convective dynamics in the underlying oceans 
\citep[e.g.][]{Soderlund20}.  We hypothesize that the broad array of 
$Nu_p/Nu_e$ solutions found in the models \citep[e.g.,][]{Soderlund19, 
Amit20, Bire22} could possibly arise because convection is not active within 
the tangent cylinder in some of the models, and is not rapidly-rotating in 
others.  Our results suggest that quantitative comparisons in heat flux 
profiles can only be made between models having similar latitudinal 
distributions of convective activity and comparable Rossby number values.

\begin{figure}
 \centering
 \includegraphics[width=\textwidth]{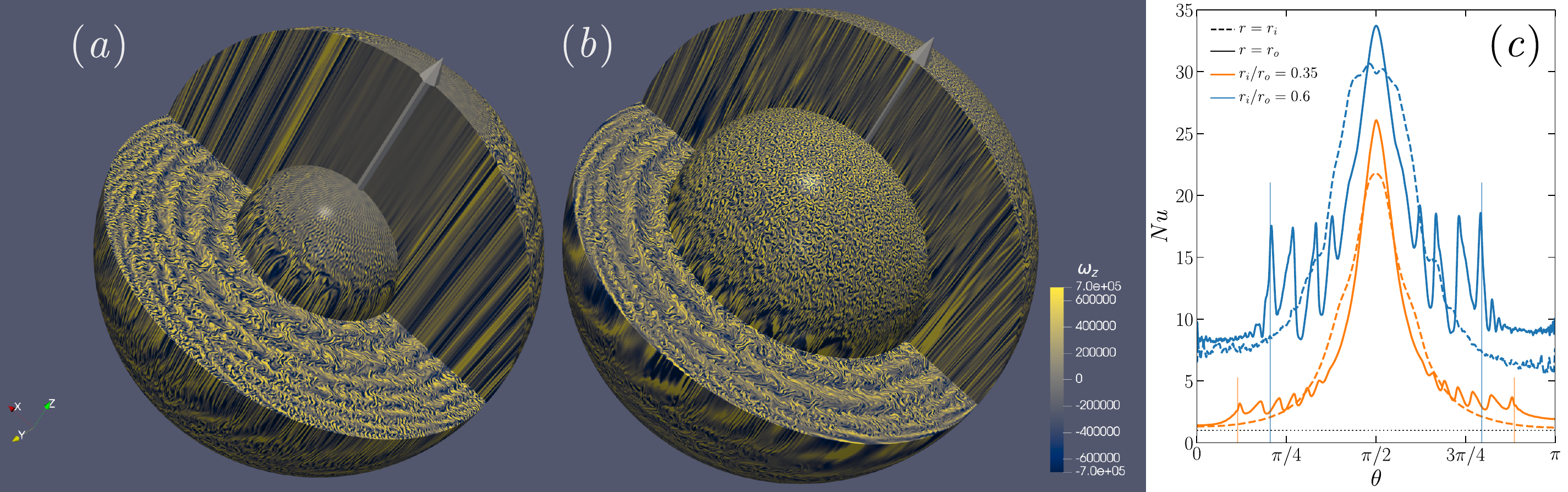}
 \caption{(\textit{a}-\textit{b}) Meridional sections, equatorial cut and 
radial surfaces of the axial 
component of the vorticity $\omega_z =\vec{e_z}\cdot 
\vec{\nabla}\times\vec{u}$. Panel (\textit{a})
corresponds to a numerical model with $r_i/r_o=0.35$, $g=r/r_o$, $E=10^{-7}$, 
$Ra=10^{11}$ and $Pr=1$, while panel (\textit{b}) corresponds 
to a numerical model with $r_i/r_o=0.6$, $g=(r_o/r)^2$, $E=3\times 10^{-7}$, 
$Ra=1.3\times 10^{10}$ and $Pr=1$. (\textit{c}) Local Nusselt number at both 
spherical shell boundaries as a function of the colatitude. The orange and 
blue lines correspond to the numerical model shown in panel (\textit{a}) and 
(\textit{b}), respectively. The location of the tangent cylinder for both 
radius ratios are marked by vertical solid lines.}
\label{fig:snaps}
\end{figure}

Establishing asymptotically-accurate trends for $Nu_p/Nu_e$ also 
requires accurate scaling laws for the equatorial heat transfer. A 
brief inspection of Fig.~\ref{fig:thRaEkNu} reveals the complexity of 
$Nu_e(\widetilde{Ra})$, and its lack of any clear power law trend. To further 
complicate this task, zonal jets tend to develop in no-slip cases with $E 
\lesssim 10^{-6}$, which can substantively alter the patterns of convective 
heat flow. Figure~\ref{fig:snaps} shows (\textit{a},\textit{b}) axial 
vorticity $\omega_z 
=\vec{e_z}\cdot \vec{\nabla}\times\vec{u}$ snapshots and (\textit{c}) 
latitudinal heat flux profiles
for two $E < 10^{-6}$ simulations with different radius ratios. Convection in 
the (\textit{a})
$r_i/r_o=0.35$ case is sub-critical inside the TC, while it is space-filling in 
the (\textit{b}) $r_i/r_o=0.6$ simulation. 
In the latter case, polar convection develops as small-scale  
axially-aligned vortices which do not drive jets within the TC. In contrast, the 
convective motions outside the TC are already sufficiently turbulent in both 
cases to trigger the formation of zonal jets. These jet flows manifest via
the formation of alternating, concentric rings of positive and negative axial 
vorticity. These coherent zonal motions act to reduce the heat transfer 
efficiency in the regions of intense shear where the zonal velocities become of 
comparable amplitude to the convective flow \citep[e.g.][]{Aurnou08, Yadav16, 
Guervilly17,Raynaud18,Soderlund19}. Thus, the outer boundary heat flux profile 
$Nu_o(\theta)$ in Fig.~\ref{fig:snaps}(\textit{c}) adopts a strongly undulatory 
structure exterior to the TC.  
The asymptotic scaling behaviour of $Nu_e$ is hence intimately related to the 
spatial distribution and amplitude of the zonal jets that develop in the shell, 
a topic for future investigations of rotating convective turbulence 
\citep[e.g.][]{Lonner22}.

\begin{acknowledgments}
We thank S.~Stellmach and K.~Julien for sharing their planar convection data. Simulations 
requiring longer time integrations to gather diagnostics 
were computed on GENCI (Grant 2021-A0070410095) and on the \texttt{S-CAPAD} platform at IPGP. 
JMA gratefully acknowledges the support of the NSF Geophysics Program (EAR 2143939). 
Lastly, we thank the University of Leiden's Lorentz Center, where this study 
was resuscitated during the ``Rotating Convection: from the Lab to the Stars'' 
workshop.

Declaration of Interests. The authors report no conflict of interest.
\end{acknowledgments}

\bibliographystyle{jfm}


\begin{thebibliography}{48}
\expandafter\ifx\csname natexlab\endcsname\relax\def\natexlab#1{#1}\fi

\bibitem[{Amit} {\em et~al.\/}(2020){Amit}, {Choblet}, {Tobie}, {Terra-Nova},
  {{\v{C}}adek} \& {Bouffard}]{Amit20}
{\sc {Amit}, H., {Choblet}, G., {Tobie}, G., {Terra-Nova}, F., {{\v{C}}adek},
  O. \& {Bouffard}, M.} 2020 {Cooling patterns in rotating thin spherical
  shells - Application to Titan's subsurface ocean}. {\em \icarus\/} {\bf 338},
  113509.

\bibitem[{Aujogue} {\em et~al.\/}(2018){Aujogue}, {Poth{\'e}rat}, {Sreenivasan}
  \& {Debray}]{Aujogue18}
{\sc {Aujogue}, K{\'e}lig, {Poth{\'e}rat}, Alban, {Sreenivasan}, Binod \&
  {Debray}, Fran{\c{c}}ois} 2018 {Experimental study of the convection in a
  rotating tangent cylinder}. {\em Journal of Fluid Mechanics\/} {\bf 843},
  355--381.

\bibitem[{Aurnou} {\em et~al.\/}(2003){Aurnou}, {Andreadis}, {Zhu} \&
  {Olson}]{Aurnou03}
{\sc {Aurnou}, Jonathan, {Andreadis}, Steven, {Zhu}, Lixin \& {Olson}, Peter}
  2003 {Experiments on convection in Earth's core tangent cylinder}. {\em Earth
  and Planetary Science Letters\/} {\bf 212}~(1-2), 119--134.

\bibitem[{Aurnou} {\em et~al.\/}(2008){Aurnou}, {Heimpel}, {Allen}, {King} \&
  {Wicht}]{Aurnou08}
{\sc {Aurnou}, J.M., {Heimpel}, M.H., {Allen}, L.A., {King}, E.M. \& {Wicht},
  J.} 2008 {Convective heat transfer and the pattern of thermal emission on the
  gas giants}. {\em Geophysical Journal International\/} {\bf 173}~(3),
  793--801.

\bibitem[{Aurnou} {\em et~al.\/}(2020){Aurnou}, {Horn} \& {Julien}]{Aurnou20}
{\sc {Aurnou}, J.M., {Horn}, S. \& {Julien}, K.} 2020 {Connections between
  nonrotating, slowly rotating, and rapidly rotating turbulent convection
  transport scalings}. {\em Physical Review Research\/} {\bf 2}~(4), 043115.

\bibitem[{Aurnou} {\em et~al.\/}(2015){Aurnou}, {Calkins}, {Cheng}, {Julien},
  {King}, {Nieves}, {Soderlund} \& {Stellmach}]{Aurnou15}
{\sc {Aurnou}, J.~M., {Calkins}, M.~A., {Cheng}, J.~S., {Julien}, K., {King},
  E.~M., {Nieves}, D., {Soderlund}, K.~M. \& {Stellmach}, S.} 2015 {Rotating
  convective turbulence in Earth and planetary cores}. {\em Physics of the
  Earth and Planetary Interiors\/} {\bf 246}, 52--71.

\bibitem[{Bire} {\em et~al.\/}(2022){Bire}, {Kang}, {Ramadhan}, {Campin} \&
  {Marshall}]{Bire22}
{\sc {Bire}, S., {Kang}, W., {Ramadhan}, A., {Campin}, J.-M. \& {Marshall}, J.}
  2022 {Exploring Ocean Circulation on Icy Moons Heated From Below}. {\em
  Journal of Geophysical Research (Planets)\/} {\bf 127}~(3), e07025.

\bibitem[{Busse} \& {Cuong}(1977)]{Busse77}
{\sc {Busse}, F.~H. \& {Cuong}, P.~G.} 1977 {Convection in rapidly rotating
  spherical fluid shells}. {\em Geophysical and Astrophysical Fluid Dynamics\/}
  {\bf 8}~(1), 17--41.

\bibitem[{Cabanes} {\em et~al.\/}(2017){Cabanes}, {Aurnou}, {Favier} \& {Le
  Bars}]{Cabanes17}
{\sc {Cabanes}, S., {Aurnou}, J.~M., {Favier}, B. \& {Le Bars}, M.} 2017 {A
  laboratory model for deep-seated jets on the gas giants}. {\em Nature
  Physics\/} {\bf 13}~(4), 387--390.

\bibitem[{Calkins}(2018)]{Calkins18}
{\sc {Calkins}, M.~A.} 2018 {Quasi-geostrophic dynamo theory}. {\em Physics of
  the Earth and Planetary Interiors\/} {\bf 276}, 182--189.

\bibitem[Cao {\em et~al.\/}(2018)Cao, Yadav \& Aurnou]{Cao18}
{\sc Cao, H., Yadav, R.~K. \& Aurnou, J.~M.} 2018 {Geomagnetic polar minima do
  not arise from steady meridional circulation}. {\em Proceedings of the
  National Academy of Sciences\/} {\bf 115}~(44), 11186--11191.

\bibitem[{Cheng} {\em et~al.\/}(2018){Cheng}, {Aurnou}, {Julien} \&
  {Kunnen}]{Cheng18}
{\sc {Cheng}, J.S., {Aurnou}, J.M., {Julien}, K. \& {Kunnen}, R.P.J.} 2018 {A
  heuristic framework for next-generation models of geostrophic convective
  turbulence}. {\em Geophysical and Astrophysical Fluid Dynamics\/} {\bf
  112}~(4), 277--300.

\bibitem[{Cheng} {\em et~al.\/}(2015){Cheng}, {Stellmach}, {Ribeiro},
  {Grannan}, {King} \& {Aurnou}]{Cheng15}
{\sc {Cheng}, J.~S., {Stellmach}, S., {Ribeiro}, A., {Grannan}, A., {King},
  E.~M. \& {Aurnou}, J.~M.} 2015 {Laboratory-numerical models of rapidly
  rotating convection in planetary cores}. {\em Geophysical Journal
  International\/} {\bf 201}, 1--17.

\bibitem[{Christensen} \& {Aubert}(2006)]{Christensen06}
{\sc {Christensen}, U.~R. \& {Aubert}, J.} 2006 {Scaling properties of
  convection-driven dynamos in rotating spherical shells and application to
  planetary magnetic fields}. {\em Geophysical Journal International\/} {\bf
  166}, 97--114.

\bibitem[{Dormy} {\em et~al.\/}(2004){Dormy}, {Soward}, {Jones}, {Jault} \&
  {Cardin}]{Dormy04}
{\sc {Dormy}, E., {Soward}, A.~M., {Jones}, C.~A., {Jault}, D. \& {Cardin}, P.}
  2004 {The onset of thermal convection in rotating spherical shells}. {\em
  Journal of Fluid Mechanics\/} {\bf 501}, 43--70.

\bibitem[{Ecke} \& {Niemela}(2014)]{Ecke14}
{\sc {Ecke}, R.~E. \& {Niemela}, J.~J.} 2014 {Heat Transport in the Geostrophic
  Regime of Rotating Rayleigh-B{\'e}nard Convection}. {\em Physical Review
  Letters\/} {\bf 113}~(11), 114301.

\bibitem[{Gastine}(2019)]{Gastine19}
{\sc {Gastine}, T.} 2019 {pizza: an open-source pseudo-spectral code for
  spherical quasi-geostrophic convection}. {\em Geophysical Journal
  International\/} {\bf 217}~(3), 1558--1576.

\bibitem[{Gastine} \& {Wicht}(2012)]{Gastine12}
{\sc {Gastine}, T. \& {Wicht}, J.} 2012 {Effects of compressibility on driving
  zonal flow in gas giants}. {\em \icarus\/} {\bf 219}, 428--442.

\bibitem[{Gastine} {\em et~al.\/}(2016){Gastine}, {Wicht} \&
  {Aubert}]{Gastine16}
{\sc {Gastine}, T., {Wicht}, J. \& {Aubert}, J.} 2016 {Scaling regimes in
  spherical shell rotating convection}. {\em Journal of Fluid Mechanics\/} {\bf
  808}, 690--732.

\bibitem[{Gastine} {\em et~al.\/}(2015){Gastine}, {Wicht} \&
  {Aurnou}]{Gastine15}
{\sc {Gastine}, T., {Wicht}, J. \& {Aurnou}, J.M.} 2015 {Turbulent
  Rayleigh-B{\'e}nard convection in spherical shells}. {\em Journal of Fluid
  Mechanics\/} {\bf 778}, 721--764.

\bibitem[{Gillet} \& {Jones}(2006)]{Gillet06}
{\sc {Gillet}, N. \& {Jones}, C.~A.} 2006 {The quasi-geostrophic model for
  rapidly rotating spherical convection outside the tangent cylinder}. {\em
  Journal of Fluid Mechanics\/} {\bf 554}, 343--369.

\bibitem[{Gilman} \& {Glatzmaier}(1981)]{Glatz1}
{\sc {Gilman}, P.~A. \& {Glatzmaier}, G.~A.} 1981 {Compressible convection in a
  rotating spherical shell - I - Anelastic equations}. {\em \apjs\/} {\bf 45},
  335--349.

\bibitem[{Grannan} {\em et~al.\/}(2022){Grannan}, {Cheng}, {Aggarwal},
  {Hawkins}, {Xu}, {Horn}, {S{\'a}nchez-{\'A}lvarez} \& {Aurnou}]{Grannan22}
{\sc {Grannan}, A.~M., {Cheng}, J.~S., {Aggarwal}, A., {Hawkins}, E.~K., {Xu},
  Y., {Horn}, S., {S{\'a}nchez-{\'A}lvarez}, J. \& {Aurnou}, J.~M.} 2022
  {Experimental pub crawl from Rayleigh{\textendash}B{\'e}nard to
  magnetostrophic convection}. {\em Journal of Fluid Mechanics\/} {\bf 939},
  R1.

\bibitem[{Guervilly} \& {Cardin}(2017)]{Guervilly17}
{\sc {Guervilly}, C. \& {Cardin}, P.} 2017 {Multiple zonal jets and convective
  heat transport barriers in a quasi-geostrophic model of planetary cores}.
  {\em Geophysical Journal International\/} {\bf 211}~(1), 455--471.

\bibitem[{Horn} \& {Shishkina}(2015)]{Horn15}
{\sc {Horn}, S. \& {Shishkina}, O.} 2015 {Toroidal and poloidal energy in
  rotating Rayleigh-B{\'e}nard convection}. {\em Journal of Fluid Mechanics\/}
  {\bf 762}, 232--255.

\bibitem[{Julien} {\em et~al.\/}(2016){Julien}, {Aurnou}, {Calkins},
  {Knobloch}, {Marti}, {Stellmach} \& {Vasil}]{Julien16}
{\sc {Julien}, K., {Aurnou}, J.~M., {Calkins}, M.~A., {Knobloch}, E., {Marti},
  P., {Stellmach}, S. \& {Vasil}, G.~M.} 2016 {A nonlinear model for
  rotationally constrained convection with Ekman pumping}. {\em Journal of
  Fluid Mechanics\/} {\bf 798}, 50--87.

\bibitem[{Julien} {\em et~al.\/}(2012{\natexlab{{\em a\/}}}){Julien},
  {Knobloch}, {Rubio} \& {Vasil}]{Julien12a}
{\sc {Julien}, K., {Knobloch}, E., {Rubio}, A.~M. \& {Vasil}, G.~M.}
  2012{\natexlab{{\em a\/}}} {Heat Transport in Low-Rossby-Number
  Rayleigh-B{\'e}nard Convection}. {\em Physical Review Letters\/} {\bf
  109}~(25), 254503.

\bibitem[{Julien} {\em et~al.\/}(2012{\natexlab{{\em b\/}}}){Julien}, {Rubio},
  {Grooms} \& {Knobloch}]{Julien12}
{\sc {Julien}, K., {Rubio}, A.~M., {Grooms}, I. \& {Knobloch}, E.}
  2012{\natexlab{{\em b\/}}} {Statistical and physical balances in low Rossby
  number Rayleigh-B{\'e}nard convection}. {\em Geophysical and Astrophysical
  Fluid Dynamics\/} {\bf 106}, 392--428.

\bibitem[{King} {\em et~al.\/}(2012){King}, {Stellmach} \& {Aurnou}]{King12}
{\sc {King}, E.~M., {Stellmach}, S. \& {Aurnou}, J.~M.} 2012 {Heat transfer by
  rapidly rotating Rayleigh-B{\'e}nard convection}. {\em Journal of Fluid
  Mechanics\/} {\bf 691}, 568--582.

\bibitem[{Kunnen}(2021)]{Kunnen21}
{\sc {Kunnen}, R.P.J.} 2021 {The geostrophic regime of rapidly rotating
  turbulent convection}. {\em Journal of Turbulence\/} {\bf 22}~(4-5),
  267–296.

\bibitem[{Long} {\em et~al.\/}(2020){Long}, {Mound}, {Davies} \&
  {Tobias}]{Long20}
{\sc {Long}, R.~S., {Mound}, J.~E., {Davies}, C.~J. \& {Tobias}, S.~M.} 2020
  {Scaling behaviour in spherical shell rotating convection with fixed-flux
  thermal boundary conditions}. {\em Journal of Fluid Mechanics\/} {\bf 889},
  A7.

\bibitem[{Lonner} {\em et~al.\/}(2022){Lonner}, {Aggarwal} \&
  {Aurnou}]{Lonner22}
{\sc {Lonner}, Taylor~L., {Aggarwal}, Ashna \& {Aurnou}, Jonathan~M.} 2022
  {Planetary Core-Style Rotating Convective Flows in Paraboloidal Laboratory
  Experiments}. {\em Journal of Geophysical Research (Planets)\/} {\bf
  127}~(10), e2022JE007356.

\bibitem[{Lu} {\em et~al.\/}(2021){Lu}, {Ding}, {Shi}, {Xia} \& {Zhong}]{Lu21}
{\sc {Lu}, H.-Y., {Ding}, G.-Y., {Shi}, J.-Q., {Xia}, K.-Q. \& {Zhong}, J.-Q.}
  2021 {Heat-transport scaling and transition in geostrophic rotating
  convection with varying aspect ratio}. {\em Physical Review Fluids\/} {\bf
  6}~(7), L071501.

\bibitem[{Miquel} {\em et~al.\/}(2018){Miquel}, {Xie}, {Featherstone}, {Julien}
  \& {Knobloch}]{Miquel18}
{\sc {Miquel}, B., {Xie}, J.-H., {Featherstone}, N., {Julien}, K. \&
  {Knobloch}, E.} 2018 {Equatorially trapped convection in a rapidly rotating
  shallow shell}. {\em Physical Review Fluids\/} {\bf 3}~(5), 053801.

\bibitem[{Niiler} \& {Bisshopp}(1965)]{Niiler65}
{\sc {Niiler}, P.~P. \& {Bisshopp}, F.~E.} 1965 {On the influence of Coriolis
  force on onset of thermal convection}. {\em Journal of Fluid Mechanics\/}
  {\bf 22}~(4), 753--761.

\bibitem[{Plumley} \& {Julien}(2019)]{Plumley19}
{\sc {Plumley}, M. \& {Julien}, K.} 2019 {Scaling Laws in Rayleigh-B{\'e}nard
  Convection}. {\em Earth and Space Science\/} {\bf 6}~(9), 1580--1592.

\bibitem[{Plumley} {\em et~al.\/}(2016){Plumley}, {Julien}, {Marti} \&
  {Stellmach}]{Plumley16}
{\sc {Plumley}, M., {Julien}, K., {Marti}, P. \& {Stellmach}, S.} 2016 {The
  effects of Ekman pumping on quasi-geostrophic Rayleigh-Benard convection}.
  {\em Journal of Fluid Mechanics\/} {\bf 803}, 51--71.

\bibitem[{Raynaud} {\em et~al.\/}(2018){Raynaud}, {Rieutord}, {Petitdemange},
  {Gastine} \& {Putigny}]{Raynaud18}
{\sc {Raynaud}, R., {Rieutord}, M., {Petitdemange}, L., {Gastine}, T. \&
  {Putigny}, B.} 2018 {Gravity darkening in late-type stars. I. The Coriolis
  effect}. {\em \aap\/} {\bf 609}, A124.

\bibitem[{Schwaiger} {\em et~al.\/}(2019){Schwaiger}, {Gastine} \&
  {Aubert}]{Schwaiger19}
{\sc {Schwaiger}, T., {Gastine}, T. \& {Aubert}, J.} 2019 {Force balance in
  numerical geodynamo simulations: a systematic study}. {\em Geophysical
  Journal International\/} {\bf 219}, S101--S114.

\bibitem[{Schwaiger} {\em et~al.\/}(2021){Schwaiger}, {Gastine} \&
  {Aubert}]{Schwaiger21}
{\sc {Schwaiger}, T., {Gastine}, T. \& {Aubert}, J.} 2021 {Relating force
  balances and flow length scales in geodynamo simulations}. {\em Geophysical
  Journal International\/} {\bf 224}~(3), 1890--1904.

\bibitem[{Soderlund}(2019)]{Soderlund19}
{\sc {Soderlund}, K.M.} 2019 {Ocean Dynamics of Outer Solar System Satellites}.
  {\em \grl\/} {\bf 46}~(15), 8700--8710.

\bibitem[{Soderlund} {\em et~al.\/}(2020){Soderlund}, {Kalousov{\'a}}, {Buffo},
  {Glein}, {Goodman}, {Mitri}, {Patterson}, {Postberg}, {Rovira-Navarro},
  {R{\"u}ckriemen}, {Saur}, {Schmidt}, {Sotin}, {Spohn}, {Tobie}, {Van Hoolst},
  {Vance} \& {Vermeersen}]{Soderlund20}
{\sc {Soderlund}, K.M., {Kalousov{\'a}}, K., {Buffo}, J.J., {Glein}, C.R.,
  {Goodman}, J.C., {Mitri}, G., {Patterson}, G.W., {Postberg}, F.,
  {Rovira-Navarro}, M., {R{\"u}ckriemen}, T., {Saur}, J., {Schmidt}, B.E.,
  {Sotin}, C., {Spohn}, T., {Tobie}, G., {Van Hoolst}, T., {Vance}, S.D. \&
  {Vermeersen}, B.} 2020 {Ice-Ocean Exchange Processes in the Jovian and
  Saturnian Satellites}. {\em \ssr\/} {\bf 216}~(5), 80.

\bibitem[{Sreenivasan} \& {Jones}(2006)]{Sreenivasan06}
{\sc {Sreenivasan}, Binod \& {Jones}, Chris~A.} 2006 {Azimuthal winds,
  convection and dynamo action in the polar regions of planetary cores}. {\em
  Geophysical and Astrophysical Fluid Dynamics\/} {\bf 100}~(4), 319--339.

\bibitem[{Stellmach} {\em et~al.\/}(2014){Stellmach}, {Lischper}, {Julien},
  {Vasil}, {Cheng}, {Ribeiro}, {King} \& {Aurnou}]{Stellmach14}
{\sc {Stellmach}, S., {Lischper}, M., {Julien}, K., {Vasil}, G., {Cheng},
  J.~S., {Ribeiro}, A., {King}, E.~M. \& {Aurnou}, J.~M.} 2014 {Approaching the
  Asymptotic Regime of Rapidly Rotating Convection: Boundary Layers versus
  Interior Dynamics}. {\em Physical Review Letters\/} {\bf 113}~(25), 254501.

\bibitem[Vallis(2017)]{Vallis17}
{\sc Vallis, G.K.} 2017 {\em Atmospheric and oceanic fluid dynamics\/}.
  Cambridge University Press.

\bibitem[{Wang} {\em et~al.\/}(2021){Wang}, {Santelli}, {Lohse}, {Verzicco} \&
  {Stevens}]{Wang21}
{\sc {Wang}, G., {Santelli}, L., {Lohse}, D., {Verzicco}, R. \& {Stevens},
  R.J.A.M.} 2021 {Diffusion-Free Scaling in Rotating Spherical
  Rayleigh-B{\'e}nard Convection}. {\em \grl\/} {\bf 48}~(20), e95017.

\bibitem[{Wicht}(2002)]{Wicht02}
{\sc {Wicht}, J.} 2002 {Inner-core conductivity in numerical dynamo
  simulations}. {\em Physics of the Earth and Planetary Interiors\/} {\bf 132},
  281--302.

\bibitem[{Yadav} {\em et~al.\/}(2016){Yadav}, {Gastine}, {Christensen},
  {Duarte} \& {Reiners}]{Yadav16}
{\sc {Yadav}, R.~K., {Gastine}, T., {Christensen}, U.~R., {Duarte}, L.~D.~V. \&
  {Reiners}, A.} 2016 {Effect of shear and magnetic field on the heat-transfer
  efficiency of convection in rotating spherical shells}. {\em Geophysical
  Journal International\/} {\bf 204}, 1120--1133.

\end{thebibliography}

\end{document}